\documentclass[figures,debug,overfull]{epl}

\newcommand{\form}{\chem{C_{12}EO_6} }
\newcommand{\formeau}{\chem{C_{12}EO_6/H_2O} }
\newcommand{\dgr}{ {\,}^{\circ} \mbox{C}}

\title{High-frequency rheological behaviour of a multiconnected lyotropic phase}
\shorttitle{Rheology of a multiconnected phase}
\author{D. Constantin\thanks{E-mail address ~: \email{dcconsta@ens-lyon.fr}} \and J.-F. Palierne \and \'E. Freyssingeas \and P. Oswald}
\shortauthor{D. Constantin \etal}
\institute{\'Ecole Normale Sup\'erieure de Lyon, Laboratoire de Physique, 46 All\'ee d'Italie, 69364 Lyon Cedex 07, France}


\begin{document}

\maketitle

\begin{abstract}
High-frequency (up to $\omega = 6 \,10^4 \un{rad/s}$) rheological measurements combined with light scattering investigations show that an isotropic and multiconnected phase of surfactant micelles exhibits a terminal relaxation time of a few $\un{\mu s}$, much smaller than in solutions of entangled wormlike micelles. This result is explained in terms of the local hexagonal order of the microscopic structure and we discuss its relevance for the understanding of dynamic behaviour in related systems, such as wormlike micelles and sponge phases.
\end{abstract}

In recent years, experimental evidence was presented as to the existence of isotropic phases consisting of connected surfactant micelles \cite{danino1,kato1,kato2}. It has been proposed that they provide an intermediate structure between entangled wormlike micelles and sponge phases \cite{porte1,drye1}. Indeed, experimental results \cite{porte1,appell1,khatory1} show that, in some ionic wormlike micellar systems, a dramatic decrease in both viscosity and relaxation time is induced by increasing the counterion concentration, feature that could be explained by the appearance of connections in the micellar network. On the theoretical side, models for the flow behaviour of these connected phases have been  developed  \cite{drye1,lequeux1}, and rheology data has been interpreted according to these models in order to characterize the appearance of connections, qualitatively \cite{narayanan1,hassan1,aitali1} or quantitatively \cite{in1}. Throughout this body of work, however, only the relaxation modes specific to polymer systems  have been considered. This approach is certainly valid in dilute phases with not too many connections, but it must fail when the density of connections becomes important and in concentrated systems, where the micelles begin to interact (sterically or otherwise). How does the system behave then and which are the relevant concepts ?

In this Letter, we try to answer these questions by investigating a concentrated and highly connected isotropic phase of a nonionic surfactant/water mixture. We argue that, in the absence of reptation (suppressed by the connections), it can be short-range order (for a concentrated system) that dominates the rheological behaviour.

We employ high-frequency rheology and dynamical light scattering (DLS) to study the isotropic phase in the \formeau lyotropic mixture,  where \form is the non-ionic sur\-fac\-tant hexa-ethylene glycol mono-n-dodecyl-ether, or \chem{CH_3(CH_2)_{11}(OCH_2CH_2)_6OH } (for the phase diagram see \cite{mitchell}). Its dynamic behaviour has already been investigated by measuring the shear viscosity \cite{strey1,darrigo1}, sound velocity and ultrasonic absorption \cite{darrigo1} as well as NMR relaxation rates \cite{burnell1}, all pointing to the presence of wormlike micelles (at least above 10 \% surfactant concentration by weight \cite{darrigo1}). In previous experiments \cite{sallen1,constantin} we have shown that, for 50 \% wt surfactant concentration, above the hexagonal mesophase, the isotropic phase has a structure consisting of surfactant cylinders that locally preserve the hexagonal order over a distance $d$ that varies from about $40 \, \un{nm}$ at $40 \dgr$ to $25 \, \un{nm}$ at $60 \dgr$. Between the cylinders there is a large number of thermally activated connections (with an estimated density $n \sim 10^{6} \, \un{\mu m^{-3}}$) \cite{constantin}.

We prepared the \formeau mixture with 50.0 \% \form weight concentration. The surfactant was purchased from Nikko Chemicals Ltd. and used without further purification. We used ultrapure water from Fluka Chemie AG. The mixture was carefully homogenized by repeatedly heating, stirring and centrifuging and then allowed to equilibrate at room temperature over a few days.

\begin{figure}
\onefigure[scale=0.8]{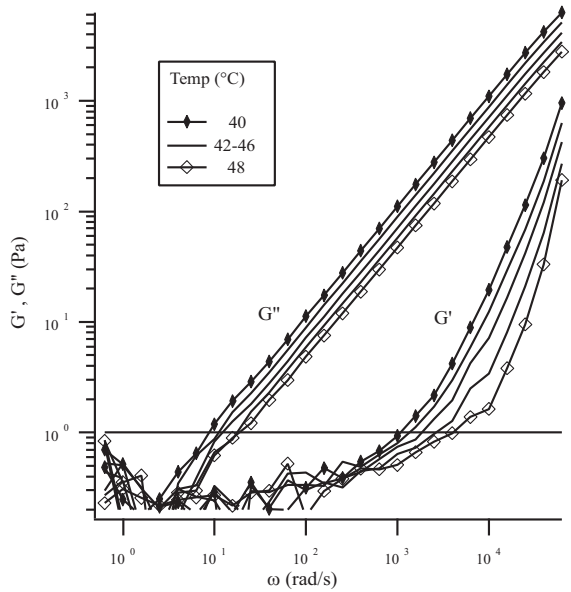}
\caption{\protect\small Storage ($G'$) and loss ($G''$) moduli as a function of $\omega$ at different temperatures. Solid curves are guides for the eye. Only values above $1 \un{Pa}$ (solid horizontal line) are relevant.}
\label{fig1}
\end{figure}

Rheology measurements were performed in a piezorheometer, the principle of which has been described in reference \cite{cagnon}~: the liquid sample of thickness $60 \un{\mu m}$ is contained between two glass plates mounted on piezoelectric ceramics. One of the plates is made to oscillate vertically with an amplitude of about 1 nm by applying a sine wave to the ceramic. This movement induces a squeezing flow in the sample and the stress transmitted to the second plate  is measured by the other piezoelectric element. The shear is extremely small~: $\gamma \leq 10^{-4}$, so the sample structure is not altered by the flow. The setup allows us to measure the storage ($G'$) and loss ($G''$) shear moduli for frequencies ranging from $1$ to $6 \, 10^{4} \un{rad/s}$ with five points per frequency decade. The entire setup is temperature regulated within $0.05 \dgr$ and hermetically sealed to avoid evaporation.

Ten temperature points in the isotropic phase have been investigated, from $38.85 \dgr$ (transition temperature from the hexagonal phase) up to $48 \dgr$. The results are displayed in figure \ref{fig1}. For clarity, only curves corresponding to 40, 42, 44, 46, and $48 \dgr$ are plotted. Values below $1 \un{Pa}$ (solid horizontal line) are not reliable, as the signal/noise ratio becomes poor. At low frequencies, the response is purely viscous; it is only above $\omega = 10^{3} \un{rad/s}$ that there is a noticeable increase in the value of the storage modulus $G'$.

\begin{figure}
\onefigure[scale=0.5]{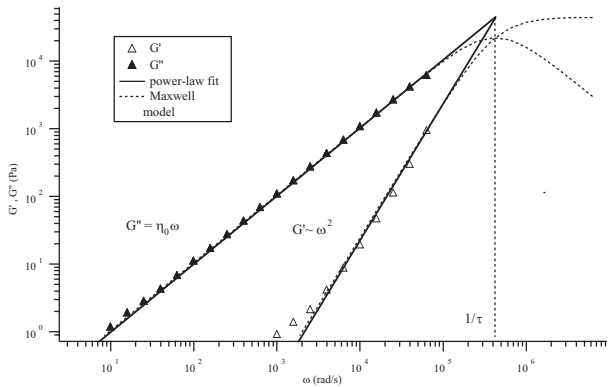}
\caption{\protect\small Typical data fit (T=$40 \dgr$)~: thick lines, power--law fit ($G'  \propto \omega ^2$ and $G''  = \eta _0 \omega$). The curves cross at a frequency $\omega=1/\tau$, where $\tau$ is the terminal relaxation time. Dotted curves, Maxwell model fit (eq. \ref{maxwell}) with the parameters $\eta _0$ and $\tau$.}
\label{fig2}
\end{figure}

On general grounds, the low-frequency behaviour of the storage and loss moduli in a fluid is \cite{ferry}~: $G'  \propto \omega ^2$ and $G'' \propto \omega$. The slope of $G'$ vs. $\omega$ yields the "zero-shear viscosity" $\eta _0$ and the two curves cross at a frequency $\omega=1/\tau$, where $\tau$ is the terminal relaxation time. The ratio $\eta _0 / \tau$ defines a shear modulus.
If $\tau$ is the only relevant time scale in the system, the complex modulus
$G^*(\omega) = G' + i G''$ has a simple analytical expression, known as the Maxwell model \cite{ferry}~:
\begin{equation}
\label{maxwell}
G^*(\omega) = \frac{i \omega \eta _0}{1+i \omega \tau} \, .
\end{equation}
The relaxation time $\tau$ separates two regimes~: for $\omega \tau \ll 1$, the system can be considered as a viscous fluid with viscosity $\eta _0$, while for $\omega \tau \gg 1$ it exhibits elasticity, with a shear modulus $G_{\infty} = \eta _0 / \tau$.

\begin{figure}
\onefigure[scale=0.8]{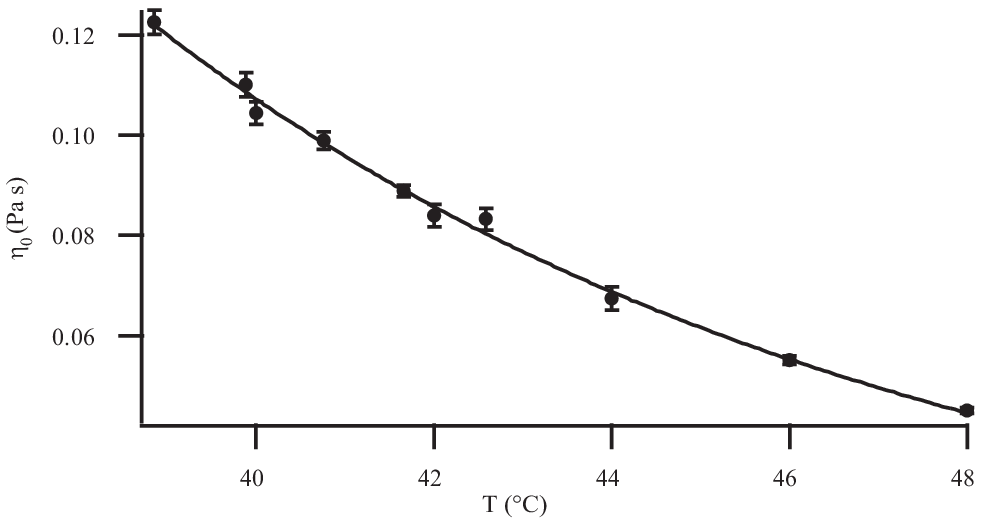}
\onefigure[scale=0.8]{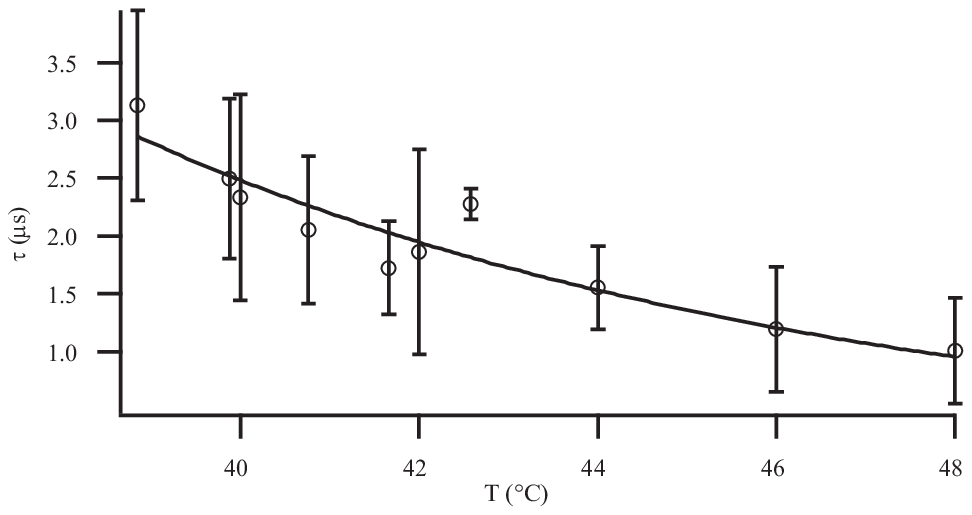}
\caption{\protect\small Static viscosity $\eta _0$ and relaxation time $\tau$ as a function of temperature. Solid line is  an Arrhenius law fit (see text). The curves start at $38.85 \dgr$, the transition temperature from the hexagonal phase.}
\label{fig3}
\end{figure}

As shown in \ref{fig2}, we obtain robust results for the static viscosity $\eta _0$ and for the relaxation time $\tau$ (plotted vs. temperature in figure \ref{fig3}).

The temperature variation of the parameters $\eta _0$ and $\tau$ can be described by Arrhenius laws; for the viscosity~:
\begin{equation}
\label{arrhenius}
\eta _0 (T) = \eta _0 (T^*) \exp \left [ \frac{E_{\eta}}{k_B} \left (\frac{1}{T}-\frac{1}{T^*}\right ) \right ] \, ,
\end{equation}
yielding an activation energy $E_{\eta} = 35 \pm 1 \, k_B T$ (solid curve in figure \ref{fig3}). For comparison, continuous shear measurements in a Couette rheometer (Haake, model RS100), give an activation energy $E_{\eta} = 31 \,  k_B T$ \cite{sallen3}. The relaxation time has an activation energy $E_{\tau} = 38 \pm 6 \, k_B T$ (solid curve in figure \ref{fig3}). Within experimental precision, $E_{\eta} = E_{\tau}$. The high-frequency elastic modulus is therefore constant in temperature~:
\begin{equation}
\label{eq:ginf}
G_{\infty} = \eta _0 / \tau = 44 \pm 6 \, 10^3 \, \un{Pa} \, .
\end{equation}

\begin{figure}
\onefigure[scale=0.8]{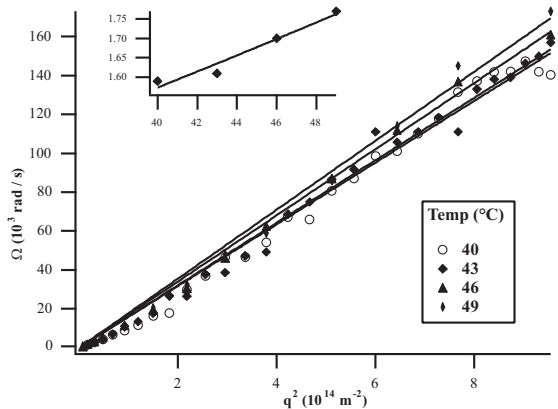}
\caption{\protect\small Relaxation frequency $\Omega$ versus $q^2$ for different temperature values. Lines are linear fits through the origin. The inset shows the values of $D$ (in units of $10^{-10}  \un{m^2/s}$) as a function of temperature. Solid curve is an Arrhenius fit.}
\label{fig5}
\end{figure}

The DLS setup uses an Ar laser ($\lambda = 514 \, \un{nm}$), delivering up to $1.5 \un{W}$, a thermostated bath of an index matching liquid (decahydronaphthalene, $n = 1.48$), a photomultiplier and a PC-controlled 256 channel Malvern correlator with sample times as fast as $0.1 \un{\mu s}$. The scattering vector $q$ varies in the range $4 \, 10^{6}$ --  $3 \, 10^{7} \un{m^{-1}}$. The signal is  monoexponential over the whole range. In figure \ref{fig5} we show  the relaxation rate $\Omega (q)$ vs. $q^2$ for temperatures between $40$ and $49 \dgr$ . The data fit well to a diffusion law (although there is a slight indication of super-diffusive behaviour). Since the scattered intensity is related to the variations in refractive index produced by concentration fluctuations, we obtain the collective diffusion constant for the concentration field; its temperature variation can be described by an Arrhenius fit (solid curve in figure \ref{fig5} -- inset) with an activation energy $E_D \simeq 4 \, k_B T$. The average value~:
\begin{equation}
D = 1.65 \, 10^{-10} \un{m^2/s}
\label{eq:diff}
\end{equation}
is in good agreement with the one previously obtained from directional-growth experiments \cite{sallen2}~: $D = 1.2 \, 10^{-10}  \un{m^2/s}$ at the transition temperature ($38.85 \dgr$).

In unconnected wormlike micellar systems \cite{drye1,cates1,cates2}, the relevant relaxation process is reptation, the micelle gradually disengaging from its initial deformed environment and adopting a stress-free configuration. The typical reptation time is given by~: $\tau _{\rm{rep}} \simeq L_{\rm{m}} ^2 / D_{\rm{c}}$, with $L_{\rm{m}}$ the average length of a micelle and $D_{\rm{c}}$ the curvilinear diffusion constant. However, if the micelles can break up (with a  lifetime $\tau _{\rm{br}}$) this provides an additional pathway for disengagement, the two resulting ends being free to recombine in a different environment. For $ \tau _{\rm{br}} \ll \tau _{\rm{rep}}$, the terminal relaxation time is given by~: $\tau = (\tau _{\rm{br}} \tau _{\rm{rep}})^{1/2}$ \cite{cates1}. As an illustration, in the CTAB/H$_2$0/KBr system  the typical micelle length is $L_{\rm{m}} \simeq 1 \mu\rm{m}$, while $\tau$ varies between 0.1 and 1 s depending on the surfactant concentration \cite{candau1}.

Let us now consider the effect of connections; following Drye and Cates \cite{drye1}, we will introduce a typical micelle length between cross-links $L_{\rm{c}}$. The effect of the connections is that reptation occurs  on distances of the order of $L_{\rm{c}}$, instead of the much larger $L_{\rm{m}}$ \cite{cates2}. This explains the fact (counterintuitive at first sight) that connecting the network does in fact reduce the viscosity. If $L_{\rm{c}}$ is small enough, the network is saturated, and the concept of entanglement is no longer applicable; neither is the reptation mechanism. The system we investigate is well in the saturated case, since the typical distance between connections on a micelle is only four times the mean distance between micelles \cite{constantin}.

\begin{figure}
\onefigure[scale=0.6]{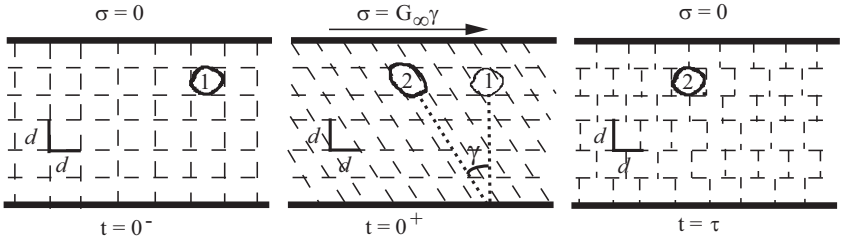}
\caption{\protect\small Schematic representation of shear in a material consisting of units of typical size $d$. One such unit (thick line contour) has been displaced between points 1 and 2. The instantaneous elastic stress is $\sigma = G_{\infty} \gamma$; it relaxes over a typical time $\tau$ given by equation \ref{tau}.}
\label{fig6}
\end{figure}

What is then the origin of viscoelasticity ? We begin the discussion of our results with the very general observation that, when a system is dynamically correlated over a typical distance $L$, one can only observe elastic behaviour by probing the system on scales smaller than the correlation distance \cite{dimension}. The time $\tau$ needed to relax the stress can then be estimated as~:
\begin{equation}
\label{tau}
 \tau  \sim L^2 / (2 \delta D) \, ,
\end{equation}
where $\delta$ is the space dimension and $D$ is the diffusion constant associated to the relaxation process (a classical example is provided by the Nabarro--Herring creep in solids \cite{quere}). The system under investigation is very concentrated so, in contrast with the semi-dilute wormlike micellar solutions usually studied, the micelle-micelle interaction plays an important role in the dynamics of the phase. This interaction locally induces hexagonal order as mentioned above; the relevant  correlation length is the distance $d$ over which the micelles preserve local order. A pictorial representation is given in figure \ref{fig6}~: consider a material with short-range order confined between two plates. The system can be seen as consisting of elasticity-endowed units of typical size $d$, the correlation distance. After applying an instantaneous shear $\gamma$ by moving the upper plate to the left, one such unit (represented in thick line) has been advected from point 1 to point 2. At time $t=0^+$ after the deformation, the stress on the upper plate is $\sigma = G_{\infty} \gamma$. Since there is no long-range restoring force, once the particles equilibrate their internal configuration (over a distance $d$), the elastic stress is completely relaxed; thus, after a time $\tau$ given by eq. \ref{tau}, $\sigma = 0$.

Does this mechanism account for the observed behaviour ? In light of the previous discussion, let us estimate the relaxation time for our system. With the value of $d$ obtained from X-ray scattering and the DLS collective diffusion coefficient (eq. \ref{eq:diff}), one has~:
\begin{equation}
\tau \simeq d^2 / (6D) \sim 10^{-6}  \un{s} \, ,
\label{tau2}
\end{equation}
in good agreement with the experimental results (figure \ref{fig3}).

A rough estimate of $G_{\infty}$ can be obtained by noticing that at short range (less than $d$), the structure of the phase resembles that of the hexagonal one, so it should exhibit a similar shear modulus when probed on very short scales.  The shear modulus of the hexagonal phase can be estimated as $G_{\rm{hex}} = k_B T / a^3 \simeq 2 \, 10^4 \un{Pa}$ (with $a=6 \un{nm}$ the lattice parameter), in agreement with our result (eq. \ref{eq:ginf}). This value can also be compared with preliminary measurements of the shear modulus in the hexagonal phase of \form \cite{pieranski1} yielding~:
\begin{equation}
G_{\rm{hex}} \simeq  2 \, 10^{5} \un{Pa}
\end{equation}
at room temperature, of the same order of magnitude as our result. The shear modulus of the hexagonal phase should vary very little with temperature, in agreement with our experimental findings.

However, our very simple model does not accurately describe the temperature variation of the physical parameters in equation \ref{tau2}. An Arrhenius fit of $d(T)$ (from the X-ray data of reference \cite{constantin}) yields an activation energy $E_d = 7 \pm 1 \,  k_B T$. From equation (7) we would expect that~:
\begin{equation}
E_{\tau} = 38 \pm 6 \, k_B T \sim 2 E_d + E_D = 18 \pm 2 \, k_B T
\end{equation}
which is clearly off by a factor of two.

A tentative explanation involves the possible anisotropy of the correlated domains; in this case, the value obtained from the X-ray diffractogram is an average between a transverse correlation length $d=d_{\bot}$ (which is the one relevant for the relaxation) and a $d_{\|}$ (which need not exhibit the same temperature variation). The same observation applies for $D$~: we measure an average value, but at small scale the structure is anisotropic.

A more detailed comparison with theory requires additional data on unsaturated structures. We are currently investigating the same isotropic phase at lower surfactant concentration, where preliminary experiments show rather complicated rheological behaviour.

Finally, we suggest that this approach can also be applied to sponge phases, the characteristic distance being $\xi$, the correlation length. These phases are equally very fluid and, at low frequency (up to at least  $10^2 \un{s^{-1}}$), display pure Newtonian behaviour \cite{snabre1,vinches1}. Within the framework of the same highly simplified model (equation \ref{tau}), we predict a relaxation time of order $\tau \sim \xi ^2 / (6D)$. For instance, in the \chem{C_{12}EO_5 }/hexanol/water system at 5.3 \% volume fraction of membrane, where
$\xi \simeq 0.1 \un{\mu m}$ and $D \simeq 2 \, 10^{-12} \un{m^2/s}$ \cite{freyssingeas1},
we expect $\tau \sim 10^{-3} \un{s}$.

In conclusion, we study the dynamics of the isotropic (micellar) phase in the \formeau mixture at high concentration, where it is highly connected. We show that the observed viscoelastic behaviour can be related to the local hexagonal order of the system.

\acknowledgments
We would like to thank P. Pieranski for communicating experimental results prior to publication and R. Strey for providing a reprint.


\begin{thebibliography}{99}
\bibitem{dimension}
Provided that the local order is 2D or 3D. If the order is lamellar (1D), the associated shear modulus is zero.
\end{thebibliography}
\end{document}